\documentclass[iop,apj]{emulateapj}
\usepackage{epsfig, color}

\newcommand{\cet}{49~Cet}
\newcommand{\boo}{$\alpha$~Boo}
\newcommand{\um}{$\mu$m}
\newcommand{\her}{\textit{Herschel}}
\newcommand{\si}{\textit{Spitzer}}
\newcommand{\lir}{$L_\mathrm{dust}/L_\star$}

\shorttitle{Gas and Dust in 49~Cet}
\shortauthors{Roberge et al.}

\begin{document}


\title{\emph{Herschel} Observations of Gas and Dust in the Unusual \object[HD9672]{49~Ceti} Debris Disk}


\author{A.~Roberge\altaffilmark{1}, I.~Kamp\altaffilmark{2}, 
B.~Montesinos\altaffilmark{3}, W.~R.~F.~Dent\altaffilmark{4}, 
G.~Meeus\altaffilmark{5}, J.~K.~Donaldson\altaffilmark{6}, J.~Olofsson\altaffilmark{7}, A.~Mo\'{o}r\altaffilmark{8},
J.-C.~Augereau\altaffilmark{9}, C.~Howard\altaffilmark{10}, C.~Eiroa\altaffilmark{5}, W.-F.~Thi\altaffilmark{9}, D.~R.~Ardila\altaffilmark{11}, G.~Sandell\altaffilmark{10}, \& P.~Woitke\altaffilmark{12}}


\altaffiltext{1}{Exoplanets \& Stellar Astrophysics Laboratory, NASA Goddard Space Flight Center, Code 667, Greenbelt, MD 20771, USA; \email{Aki.Roberge@nasa.gov}}

\altaffiltext{2}{Kapteyn Astronomical Institute, University of Groningen, 9700 AV Groningen, Netherlands}

\altaffiltext{3}{Departamento de Astrof\'{i}sica, Centro de Astrobiolog\'{i}a (INTA-CSIC), ESAC Campus, PO Box 78, 28691 Villanueva de la Ca\~{n}ada, Madrid,
Spain}
\altaffiltext{4}{ALMA/SCO, Alonso de Cordova 3107, Vitacura, Santiago, Chile}
\altaffiltext{5}{Departmento F\'{i}sica Te\'{o}rica, Facultad de Ciencias, 
Universidad Aut\'{o}noma de Madrid, Cantoblanco, 28049 Madrid, Spain}
\altaffiltext{6}{Department of Astronomy, University of Maryland, College Park, MD 20742, USA}
\altaffiltext{7}{Max Planck Institute for Astronomy, K\"{o}nigstuhl 17, D-69117, 
Heidelberg, Germany}
\altaffiltext{8}{Konkoly Observatory of the Hungarian Academy of Sciences, P.O.\ Box 67, H-1525 Budapest, Hungary}
\altaffiltext{9}{UJF-Grenoble 1/CNRS-INSU, Institut de Plan\'{e}tologie et d'Astrophysique de Grenoble, UMR 5274, 38041, Grenoble, France}

\altaffiltext{10}{SOFIA-USRA, NASA Ames Research Center, Building N232, PO~Box~1, Moffett Field, CA 94035, USA}
\altaffiltext{11}{NASA Herschel Science Center, California Institute of Technology, 1200 E.~California Blvd., Mail Stop 220-6, Pasadena, CA 91125, USA}
\altaffiltext{12}{University of Vienna, Department of Astronomy, T\"{u}rkenschanzstr.~17, 1180, Vienna, Austria}



\begin{abstract}
We present far-IR/sub-mm imaging and spectroscopy of 49~Ceti, an unusual circumstellar disk around a nearby young A1V star.  The system is famous for showing the dust properties of a debris disk, but the gas properties of a low-mass protoplanetary disk. The data were acquired with the \textit{Herschel Space Observatory} PACS and SPIRE instruments, largely as part of the ``Gas in Protoplanetary Systems'' (GASPS) Open Time Key Programme. Disk dust emission is detected in images at 70, 160, 250, 350, and 500~$\mu$m; 49~Cet is significantly extended in the 70~$\mu$m image, spatially resolving the outer dust disk for the first time. Spectra covering small wavelength ranges centered on eight atomic and molecular emission lines were obtained, including [\ion{O}{1}] 63~$\mu$m and [\ion{C}{2}] 158~\um. The \ion{C}{2} line was detected at the $5\sigma$ level -- the first detection of atomic emission from the disk. No other emission lines were seen, despite the fact that the \ion{O}{1} line is the brightest one observed in \her\ protoplanetary disk spectra \citep[e.g.][]{Meeus:2012, Dent:2013}. We present an estimate of the amount of circumstellar atomic gas implied by the \ion{C}{2} emission. The new far-IR/sub-mm data fills in a large gap in the previous spectral energy distribution (SED) of 49~Cet. A simple model of the new SED confirms the two-component structure of the disk: warm inner dust and cold outer dust that produces most of the observed excess. Finally, we discuss preliminary thermochemical modeling of the 49~Cet gas/dust disk and our attempts to match several observational results simultaneously. Although we are not yet successful in doing so, our investigations shed light on the evolutionary status of the 49~Cet gas, which might not be primordial gas but rather secondary gas coming from comets.
\end{abstract}






\section{Introduction \label{sec:intro} }

Protoplanetary disks begin as massive, gas-rich remnants of their parent interstellar molecular clouds, but rapidly evolve into tenuous, dusty disks within about 10~Myr.
These disks, called debris disks, differ from protoplanetary disks in that they are composed of secondary material recently generated by collisions between and evaporation of asteroids and comets.
Planetesimal collisions presumably give rise to rocky planets, and therefore the younger debris disks ($\lesssim 100$~Myr) provide valuable insight into the later stages of terrestrial planet formation.

Debris disks are gas-poor, as evidenced by a general lack of sub-mm carbon monoxide emission \citep[e.g.][]{Zuckerman:1995}.
However, small amounts of gas have been found in several debris disks, the most famous of which is $\beta$~Pictoris \citep[e.g.][]{Lagrange:1998,Roberge:2006}. This primarily atomic gas must also be recently produced secondary material, as many of the species seen have short lifetimes in optically-thin environments (for example, the short photoionization lifetime of \ion{C}{1}).
As for the dust, the ultimate source of the gas is the destruction of planetesimals (see \citealp{Roberge:2011} for a fuller discussion).

The well-known 49~Ceti system consists of a bright debris disk surrounding a single A1V star at a \emph{Hipparcos} distance of $59 \pm 1$~pc \citep{vanLeeuwen:2007}.
The stellar age estimated by comparison of the star's position on the H-R diagram to theoretical stellar evolutionary tracks is either $8.9^{+6.1}_{-2.4}$~Myr from pre-main sequence tracks or $61^{+119}_{-46}$~Myr from post-main sequence tracks \citep{Montesinos:2009}.
Recently, \citet{Zuckerman:2012} identified 49~Cet as a co-moving member of the $\sim 40$~Myr-old Argus Association \citep{Torres:2008}.
Spectral energy distribution (SED) fitting suggested that the \cet\ disk has two distinct components, a cold outer disk and a warmer inner one \citep[e.g.][]{Wahhaj:2007}.  The inner dust has been spatially resolved in mid-IR thermal emission but no emission features associated with \um-sized grains are observed in the \si\ IRS mid-IR spectrum of the system \citep{Wahhaj:2007,Chen:2006}. 

\cet\ is one of two debris disks that show sub-mm CO emission \citep{Zuckerman:1995, Dent:2005, Hughes:2008}; the other is HD21997, whose CO emission was recently discovered \citep{Moor:2011}. 
The CO emission from \cet\ led to questions about its evolutionary status, possibly marking it as a rare late-stage protoplanetary disk that has dissipated most but not all of its primordial gas \citep[e.g.][]{Hughes:2008}. 
Moderate gas drag at such a time may be important for establishing the final orbital characteristics of any terrestrial planets \citep[e.g.][]{Nagasawa:2005}. 

However, the relatively low fractional infrared luminosity of \cet\ \citep[$L_\mathrm{IR} / L_\star \sim 10^{-3}$;][]{Jura:1998} shows that the dust is optically thin, which is characteristic of a debris disk. 
On the strength of the system's relatively advanced $\sim 40$~Myr age, \citet{Zuckerman:2012} proposed that the observed CO gas is not primordial but is rather coming from a massive belt of colliding comet-like planetesimals.
At this time, however, the evolutionary status of 49~Cet's gas component is uncertain.

In this paper, we present far-IR/sub-mm imaging and spectroscopy of \cet\ obtained with the \textit{Herschel Space Observatory} \citep{Pilbratt:2010}.
In Section 2, the observations and data reduction are described.
Analysis and results appear in Section~3: analysis of the spatially resolved 70~\um\ image in Section~\ref{sub:image}, all photometry in Section~\ref{sub:phot}, spectroscopic analysis in Section~\ref{sub:spec}, and simple gas mass estimates in Section~\ref{sub:gas_mass}.
In Section~4, we show the new \cet\ SED, present an improved model for the stellar spectrum, and describe our simple dust model fitting. 
Preliminary modeling of the gas component is discussed in Section~5.
An examination of the evolutionary status of the \cet\ disk material appears in Section~6.
Finally, Section~7 summarizes our primary findings and some planned future work.

\section{Observations and Data Reduction}

49~Ceti was observed with the far-IR PACS instrument \citep{Poglitsch:2010}, as part of the ``Gas in Protoplanetary Systems'' (GASPS) Open Time Key Programme \citep[e.g.][]{Mathews:2010, Dent:2013}.
Single PACS scan maps at 70~\um\ and 160~\um\ (OBSID 1342188485) were obtained on 2009-12-23 using the medium scan speed ($20\arcsec \ \mathrm{s}^{-1}$).  The maps consisted of 8 scan legs with $3\arcmin$ lengths and $5\arcsec$ cross-scan steps.  The on-source exposure time for each map was 72~sec.

We also obtained three PACS spectroscopy observations in chop-nod mode.  The first was a RangeSpec observation that included 6 small wavelength ranges centered on the following transitions:  CO 73~\um, H$_2$O 79~\um, CO 90~\um, [\ion{O}{1}] 145~\um, [\ion{C}{2}] 158~\um, H$_2$O 180~\um\ (taken 2009-12-22, OBSID 1342188423, 251~sec on-source time per range).  The second was a LineSpec observation that covered two small wavelength ranges centered on the [\ion{O}{1}] 63~\um\ and DCO$^+$ 190~\um\ transitions (taken 2009-12-22, OBSID 1342188424, 370~sec on-source time per range).  The third was a deeper RangeSpec observation focused on the [\ion{C}{2}] 158~\um\ line (taken 2011-07-08, OBSID 1342223790, 670~sec on-source time).

All PACS data were calibrated with HIPE v8.2, using pipeline calibration scripts \citep{Ott:2010}.
Final scan maps for each wavelength were generated with two different pixel scales: one with $1\arcsec$ pixels for use in resolved imaging analysis (Section~\ref{sub:image}) and one with the native pixel scale of the PACS detectors ($3\farcs2$ at 70~\um\ and $6\farcs4$ at 160~\um) for use in photometry analysis (Section~\ref{sub:phot}). Since \cet\ is bright at far-IR wavelengths, standard calibration of scan maps will lead to over-subtraction of the background during high-pass filtering. Therefore, a region around the source was masked before filtering.
The absolute calibration uncertainties for the 70 and 160~\um\ scan maps are 2.64\% and 4.15\%, respectively.\footnote{PICC-ME-TN-037: \url{http://herschel.esac.esa.int/twiki/pub\newline/Public/PacsCalibrationWeb/pacs\_bolo\_fluxcal\_report\_v1.pdf} \label{foot:picc}}

PACS is an integral field spectrometer (IFS) with a $5 \times 5$ array of spectral pixels (spaxels), each $9\farcs4 \times 9\farcs4$ in size. All spectroscopy was obtained in ChopNod mode to remove telescope and sky background emission. For each spectroscopic observation, 25 \mbox{1-D} spectra were extracted, one from each spaxel. The pipeline calibration script applied a flux correction to each spectrum, accounting for the expected fraction of the point-spread function (PSF) falling outside the spaxel.  Spectra with two different pixel scales were produced: one corresponding to the native resolution of the instrument (2 pixels per resolution element, i.e.\ Nyquist sampling) and one with smaller pixels (oversampled with 3 pixels per resolution element).  All analysis was performed on the Nyquist-sampled data, which gave results with the highest signal-to-noise ($S/N$), as expected.  The oversampled spectra were used only to check results and for display purposes.

In addition to the PACS data, we made use of sub-mm \cet\ observations taken with the SPIRE instrument \citep{Griffin:2010}.
The SPIRE small scan maps at 250~\um, 350~\um, and 500~\um\ (OBSID 1342236226) were acquired in January 2012 as part of a \her\ Open Time program (OT1\_pabraham\_2, PI: P.~\'{A}brah\'{a}m).
The data were calibrated using HIPE~v8.2 and the standard pipeline calibration script, producing maps with units of Jy~beam$^{-1}$, which we converted to Jy~pixel$^{-1}$ before analysis. 
The map pixel scales were $6\arcsec$, $10\arcsec$, and $14\arcsec$ at 250~\um, 350~\um, and 500~\um, respectively.
The estimated calibration uncertainty for the SPIRE images is  7\%.\footnote{SPIRE Observer's Manual v2.4 (June~2011):\newline
\url{http://herschel.esac.esa.int/Docs/SPIRE/html/spire\_om.html} \label{foot:som}}

Finally, we re-calibrated the low-resolution \si\ IRS spectrum of \cet\ published in \citet{Chen:2006} using the FEPS \textit{Spitzer} Legacy Project pipeline.
As is typical, there was a flux disjoint between the short-low and long-low portions of the spectrum.  
We shifted the long-low portion to match the \si\ MIPS 24~\um\ flux (discussed in Section~\ref{sub:phot}), which removed the disjoint between the two halves of the spectrum.

\section{Analysis and Results \label{sec:analysis} }

\subsection{Imaging \label{sub:image}}

\begin{figure*} \centering
\epsfig{file=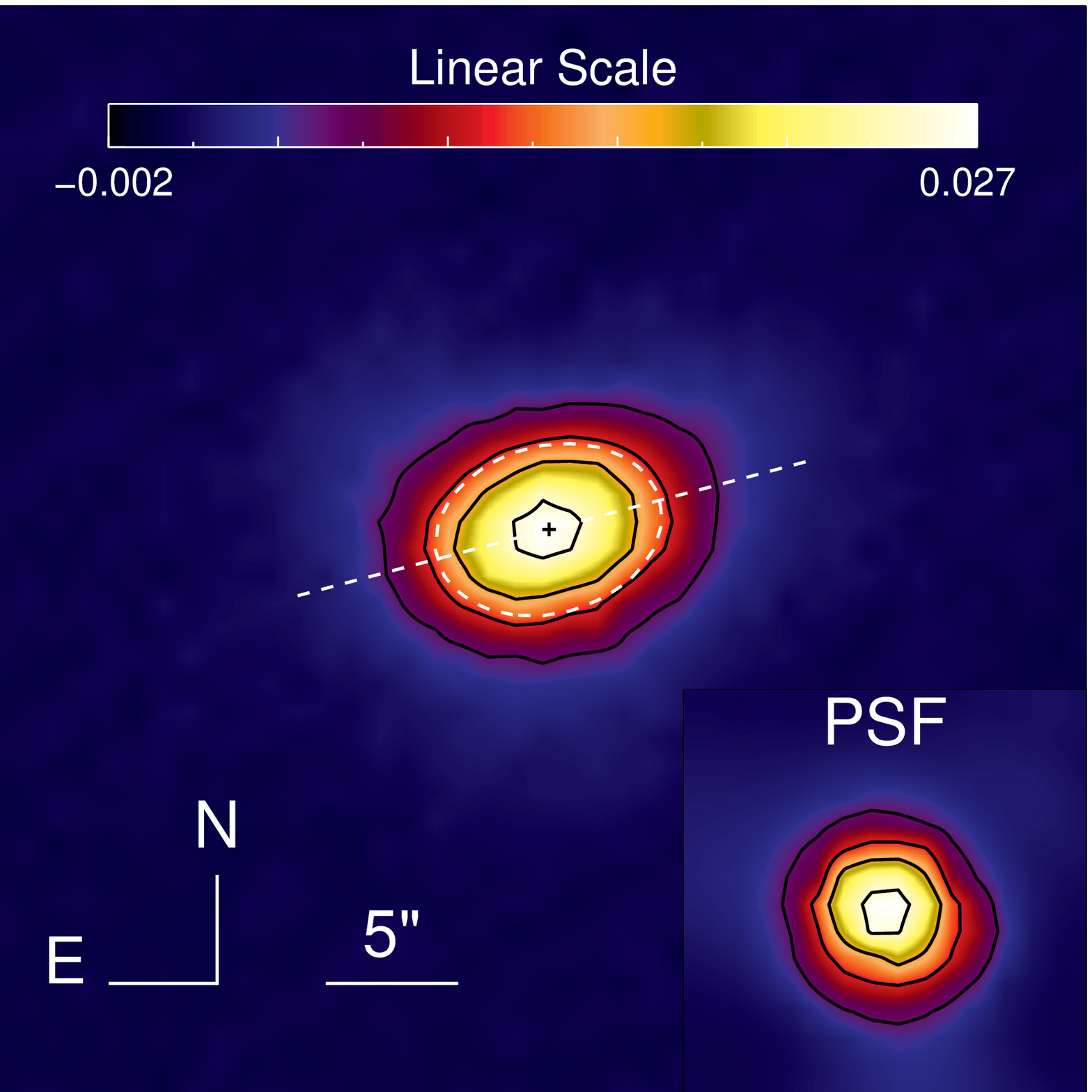, width=3.in} \hspace{0.1in}
\epsfig{file=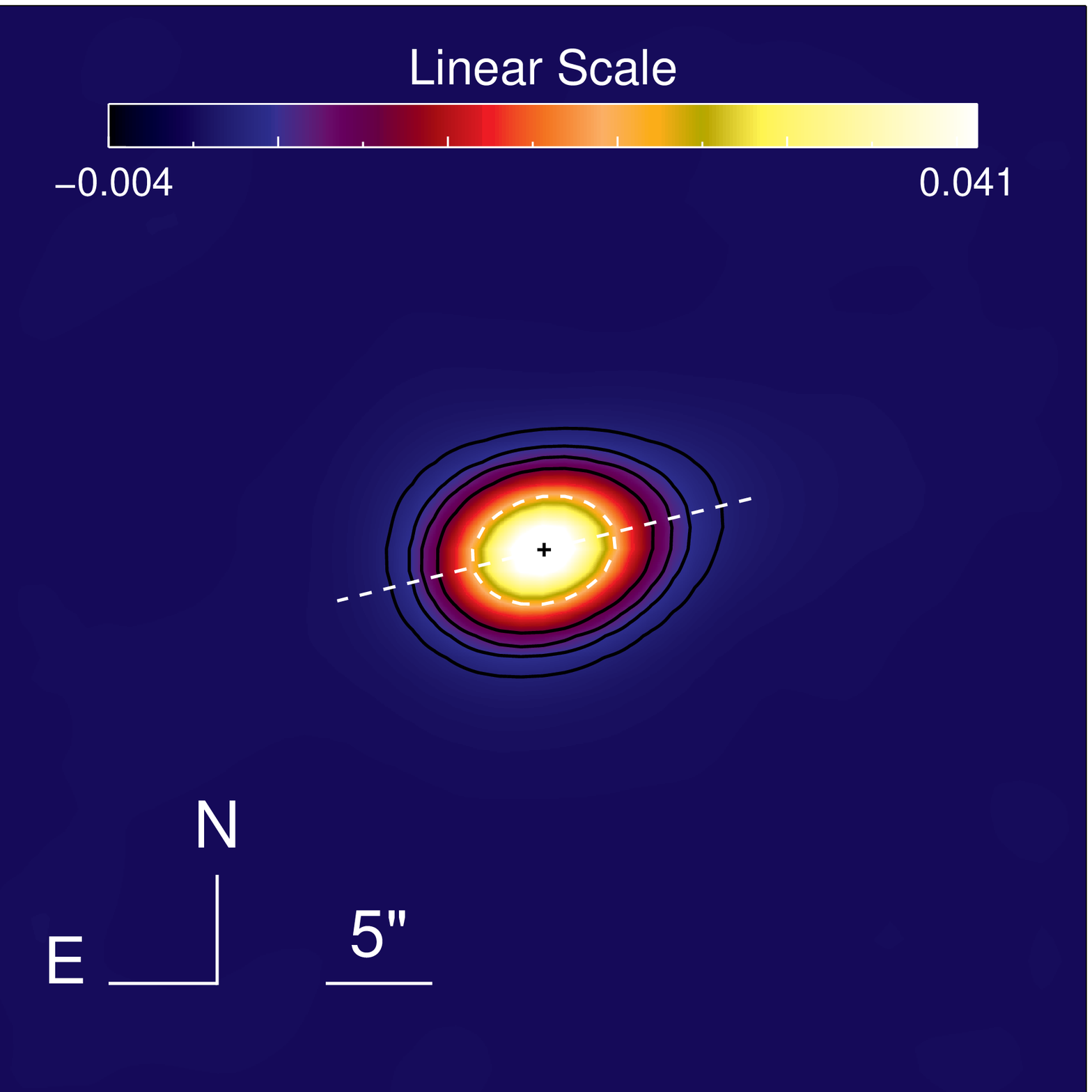, width=3.in}
\caption{\her\ PACS 70~\um\ images of 49~Cet. In both panels, the pixel scale is $1\arcsec$.  Brightness contours at 3, 6, 9, and 14 times the background noise are overlaid (solid black contours).  The dashed white ellipse shows the half-maximum contour for the best-fitting \mbox{2-D} Gaussian. The center of the best-fitting \mbox{2-D} Gaussian is marked with a black plus. 
{\bf Left:} Original calibrated scan map, with a 70~\um\ scan map of a PSF reference star (\boo) in the lower right corner.  The position angle (PA) of the \cet\ disk major axis ($-75^\circ$ E of N) is indicated with the straight dashed line.
{\bf Right:} Deconvolved scan map. The new PA of the disk major axis ($-76^\circ$ E of N) is indicated with the straight dashed line. \label{fig:cet_70}}
\end{figure*}

\cet\ is significantly extended in the 70~\um\ image compared to a similar image of the diskless PSF reference star \boo\ obtained using the same medium scan speed, as shown in the left-hand panel of Figure~\ref{fig:cet_70}. This is the first time thermal emission from dust in the outer disk of \cet\ has been spatially resolved. 
We began our analysis by fitting \mbox{2-D} Gaussians to the \cet\ and \boo\ images; the results appear in Table~\ref{tab:resolve}.
We estimated the uncertainty in our full-width at half-maximum (FWHM) measurements by performing the same analysis on a different 70~\um\ scan map of \boo.
The FWHM of the best-fitting \mbox{2-D} Gaussian varied by $0\farcs3$ between the two \boo\ images.
Adopting this value as our FWHM uncertainty, \cet\ is resolved along the major axis at the $10\sigma$ level.
The measured position angle (PA) of the disk major axis ($-75^\circ \pm 1^\circ$ E of N) is in reasonable agreement with the PA determined from spatially resolved sub-mm CO contours \citep[$-79^\circ$ E of N;][]{Hughes:2008}.
 
To estimate the true extent of the dust disk, we deconvolved the \cet\ image using the \boo\ image as the reference PSF, rotated to match the roll angle of the telescope at the time of the \cet\ observation.
The stellar contribution was removed from the \cet\ image before deconvolution, by subtracting the PSF reference image scaled so that the total flux from \boo\ matched the expected stellar flux from \cet.  Since the expected \cet\ stellar flux is only $0.2\%$ of the actual measured flux (see Section~\ref{sub:star}), this had a negligible effect on the deconvolution. We employed the Richardson-Lucy deconvolution method, conservatively limiting the number of iterations to five. The reason for choosing this number will become apparent later.
The resulting deconvolved image appears in the right-hand panel of Figure~\ref{fig:cet_70}.

The deconvolved disk image shows no sign of a central clearing or any asymmetric structure. 
The position angle of the major axis changed only slightly.
We re-fit a \mbox{2-D} Gaussian to the deconvolved \cet\ image; the results are included in Table~\ref{tab:resolve}.
The half-width at half maximum (HWHM) along the major axis is $\sim 200$~AU.
In their modeling of the \cet\ gas and dust disk, \citet{Hughes:2008} found that they had to set the outer radius of the disk to 200~AU.
While the agreement between that value and our HWHM is partly coincidental, it does indicate that the cold CO gas observed in the sub-mm and the cold dust we observed here may be co-spatial.

During the deconvolution procedure, we found that we could make \cet\ significantly smaller than \boo\ along the minor axis by iterating many times.  But no matter how many iterations were used (up to $\sim 1000$), \cet\ was always larger than \boo\ along the major axis. We took this as an indication that the disk is  significantly resolved along the major axis but not the minor axis.  Therefore, we chose the number of deconvolution iterations such that the FWHM of the disk along the minor axis nearly matched that of \boo. This was accomplished in 5 iterations. Since the minor axis is not resolved, we only obtain a lower limit on the outer disk inclination of $i \geq 44^\circ$, assuming the disk is inherently circular. This is broadly consistent with the high inclination derived from the sub-mm CO contours \citep[$i = 90^\circ \pm 5^\circ$;][]{Hughes:2008}.
\cet\ is not significantly extended in the 160~\um\ image or the SPIRE images.

\begin{deluxetable}{lccc}
\tablewidth{0pt}
\tablecolumns{4}
\tablecaption{\cet\ Disk Geometry from the 70~\um\ Image \label{tab:resolve} }
\tablehead{ \colhead{ } & \colhead{49~Cet}  & \colhead{$\alpha$~Boo} & \colhead{49~Cet Deconvol.} }
\startdata
Major axis PA (E of N)    & $-75^\circ \pm 1^\circ$ &  & $-76.1^\circ \pm 0.5^\circ$\\
Major axis (FWHM)         & $8\farcs4$  & $5\farcs3$  & $6\farcs7$ \\
Semi-major axis (HWHM)    & 250 AU      &             & 200 AU \\
Minor axis (FWHM)         & $6\farcs1$  & $4\farcs9$  & $4\farcs8$ \\
Semi-minor axis (HWHM)    & $\leq 180$ AU      &         & $\leq 140$ AU \\
Inclination               & $\geq 44^\circ$   &       & $\geq 44^\circ$ 
\enddata
\tablecomments{Estimated uncertainty of a FWHM measurement is $0\farcs3 = 18$~AU.} 
\end{deluxetable}

\subsection{Photometry \label{sub:phot}}

To measure the total flux from \cet\ at 70 and 160~\um, aperture photometry was performed on the scan maps with the native pixel scale. 
The radii of the circular object apertures were $14\arcsec$ and $22\arcsec$ at 70 and 160~\um, respectively.
These are larger than recommended by the \her\ PACS Instrument Control Centre (ICC) for aperture photometry of unresolved faint sources, since \cet\ is bright and the disk is spatially resolved at 70~\um.  The apertures were chosen to be roughly 1.5 times the FWHM of the disk at that wavelength (Section~\ref{sub:image}), so that they encompass nearly all of the disk flux.
Background subtraction was performed using the mean brightness in sky annuli 
$20\arcsec - 26\arcsec$ from the star center for the 70~\um\ image and $28\arcsec - 34\arcsec$ for the 160~\um\ image.
Aperture corrections provided by the \her\ PACS ICC\footnotemark[13] were applied to the total fluxes, but color corrections were not applied.

Assuming background-limited imaging, the statistical uncertainties in the final PACS fluxes are given by 
\begin{equation}
\sigma_\mathrm{stat} = \frac{ \sigma_\mathrm{rms} } {\alpha_\mathrm{cor} \; x_\mathrm{cor} }
\sqrt{ n_\mathrm{ap} \left( 1 + \frac{ n_\mathrm{ap} }{ n_\mathrm{sky} } \right) } \; ,
\end{equation}
where $\sigma_\mathrm{rms}$ is the standard deviation of the pixels in the sky annulus, $\alpha_\mathrm{cor}$ is the aperture correction, $x_\mathrm{cor}$ is the correlated noise correction\footnotemark[13] (0.95 at 70~\um, 0.88 at 160~\um), $n_\mathrm{ap}$ is the number of pixels in the object aperture, and $n_\mathrm{sky}$ is the number of pixels in the sky annulus.
The absolute calibration uncertainties (2.64\% at 70~\um, 4.15\% at 160~\um) were added in quadrature to the statistical uncertainties to give the total flux uncertainties.
The total \cet\ fluxes at 70 and 160~\um\ are $F_\mathrm{70~\mu m} = 2.142 \ \mathrm{Jy} \pm 0.058 \ \mathrm{Jy}$ and $F_\mathrm{160~\mu m} = 1.004 \ \mathrm{Jy} \pm 0.053 \ \mathrm{Jy}$.

We performed aperture photometry on the SPIRE maps using circular apertures with radii of $22\arcsec$, $30\arcsec$, and $42\arcsec$, respectively.
For each wavelength, the background was estimated in a sky annulus $60\arcsec - 90\arcsec$ from the star center.
Aperture corrections were applied and the fluxes color-corrected for a $F_{\nu} \propto \nu^2$ point-source spectrum.\footnote{SPIRE Data Reduction Guide v2.0, (March~2012):\newline \url{http://herschel.esac.esa.int/\mbox{hcss-doc-8.0}/load/spire\_drg/html\newline /spire\_drg.html}}
The total \cet\ fluxes at 250, 350, and 500~\um\ are 
$F_\mathrm{250~\mu m} = 0.372 \ \mathrm{Jy} \pm 0.027 \ \mathrm{Jy}$,
$F_\mathrm{350~\mu m} = 0.180 \ \mathrm{Jy} \pm 0.014 \ \mathrm{Jy}$, and
$F_\mathrm{500~\mu m} = 0.086 \ \mathrm{Jy} \pm 0.009 \ \mathrm{Jy}$.
The flux uncertainties include statistical and calibration uncertainties; since \cet\ is relatively bright, the latter dominate.

For use in the \cet\ SED and to correct the \si\ IRS long-low spectrum, we also calculated the continuum flux at 24~\um\ from archival \textit{Spitzer} MIPS imaging taken as part of the MIPS-GTO program DISKLEGACY (PI: G.~Rieke, AOR\# 21942016).
The post-BCD image was downloaded from the Spitzer Heritage Archive and aperture photometry performed using the aperture size and correction from \citet{Su:2006}.
We found $F_\mathrm{24~\mu m} = 0.259 \ \mathrm{Jy} \pm 0.010 \ \mathrm{Jy}$, where the error is given by the $4\%$ absolute calibration uncertainty\footnote{MIPS Instrument Handbook v3.0 (March 2011):\newline \url{http://irsa.ipac.caltech.edu/data/SPITZER/docs/mips/\newline mipsinstrumenthandbook}}  (the statistical uncertainty is negligible).

\subsection{Spectroscopy \label{sub:spec}}

In all the \cet\ PACS spectra, significant continuum or line emission appears only in the central spaxel.  We verified that the star was well-centered on the array during the spectroscopic observations (shifts $\leq 0\farcs5 \leq 0.05$~spaxel).  This was done by taking theoretical PSFs, offsetting them on a virtual PACS IFS, and comparing the fraction of flux in each spaxel as a function of shift to the observed values. All further analysis discussed here was performed on the spectra from the central spaxel, using the Nyquist-sampled data.  Only one emission line is detected: [\ion{C}{2}] 158~\um\ (discussed below).  

The spectra without significant emission lines were analyzed in the following manner. For each spectrum, we least-squares fit a 1st-degree polynomial to the continuum. The statistical flux uncertainties were then estimated by taking the standard deviation of the fluxes minus the continuum fit, in a range centered on the expected line position. The width of the range was chosen so that roughly $68\%$ of the pixels were within $1\sigma$ of the continuum. The continuum flux values at the expected line center wavelengths appear in Table~\ref{tab:spec}.  The flux errors given are the statistical flux uncertainties and the absolute flux calibration uncertainties\footnote{PACS Observer's Manual v2.4 (Dec.~2011): \url{http://herschel.esac.esa.int/Docs/PACS/html/pacs\_om.html}} (11\% at 63~\um, 12\% at other wavelengths) added in quadrature. 

\begin{figure*} \centering
\epsfig{file=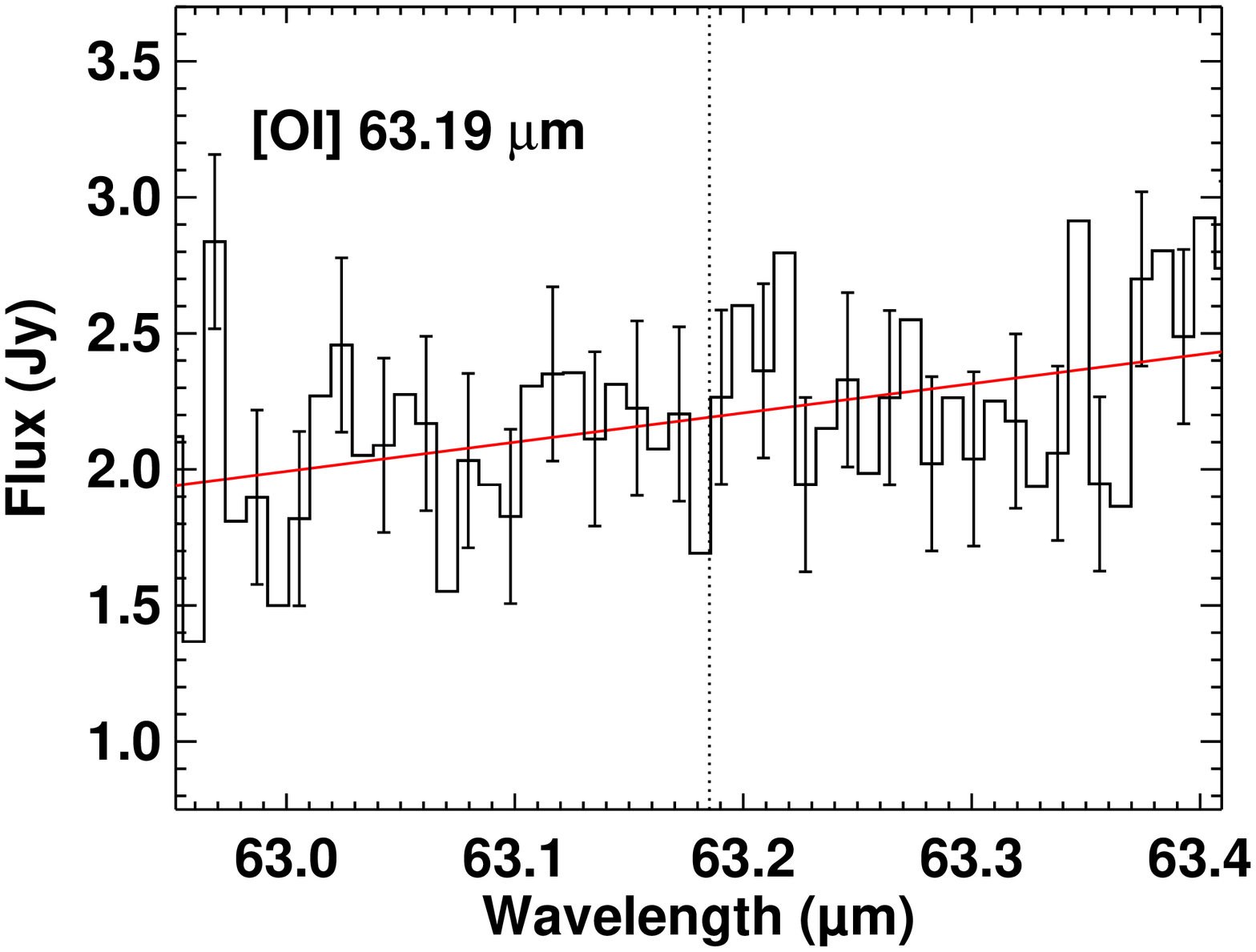, width=3.25in} \hspace{0.1in}
\epsfig{file=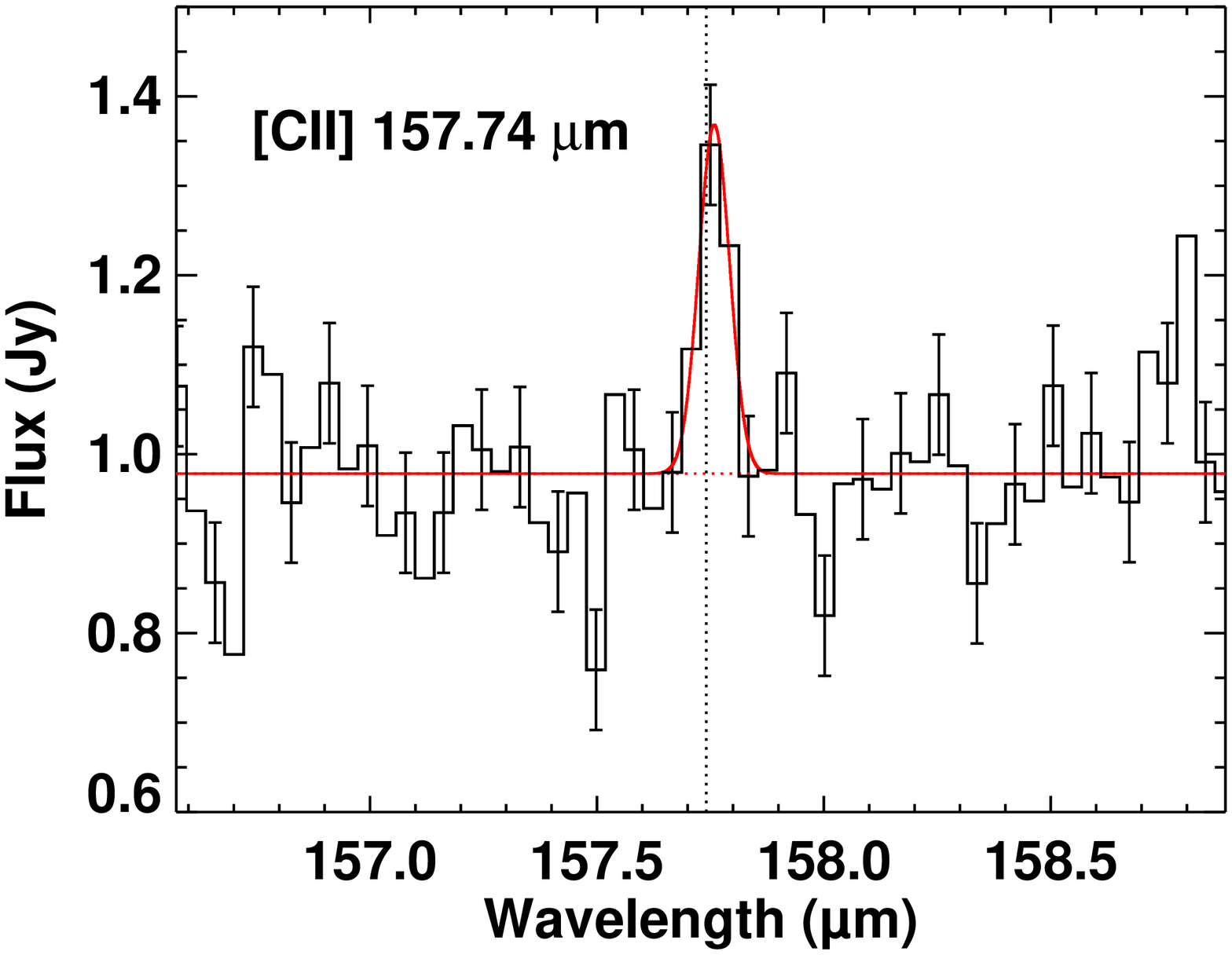, width=3.25in}
\caption{\cet\ spectra showing the regions around the [\ion{O}{1}] 63~\um\ line {\bf (left)} and the [\ion{C}{2}] 158~\um\ line {\bf (right)}.  The rest wavelengths of the lines are marked with dotted lines.  No significant \ion{O}{1} emission is seen, while the \ion{C}{2} line is detected at the $\sim 5\sigma$ level. Our best-fitting models for the continuum or the continuum plus emission line are over-plotted with red solid lines.  The data shown here have the oversampled wavelength scale, highlighting the shape of the [\ion{C}{2}] 158~\um\ emission line, which is unresolved.  \label{fig:spec} }
\end{figure*}

Upper limits on the total emission line fluxes were calculated by integrating the continuum-subtracted spectrum over a small wavelength range centered on the expected line wavelength and propagating the final statistical flux errors; the wavelength range was $\pm 1.5 \times$ the expected width of an unresolved emission line. The results appear in Table~\ref{tab:spec}.  A plot of the region around the undetected [\ion{O}{1}] 63~\um\ line appears in the left-hand panel of Figure~\ref{fig:spec}; this is the brightest line observed from protoplanetary disks with \her\ \citep[e.g.][]{Meeus:2012, Dent:2013}.

The only line detected, [\ion{C}{2}] 158~\um, was seen in the original shallow spectrum, though it was not a significant detection in the data calibrated with earlier versions of HIPE (e.g.\ v4.2). Therefore, we obtained a deeper follow-up spectrum to confirm the detection, shown in the right-hand panel of Figure~\ref{fig:spec}. 
For both the shallow and deep spectra, we analyzed the [\ion{C}{2}] 158~\um\ line in the following way.  We modeled the continuum plus line by least-squares fitting the sum of a 1st-degree polynomial and a Gaussian.  Then the statistical flux uncertainties were estimated by taking the standard deviation of the fluxes minus the best-fitting total model, in a range around the line.  This time, the width of the range was chosen to maximize the $S/N$ of the emission line ($\pm 7.5 \times$ the instrumental FWHM). 
As before, the total \ion{C}{2} emission line flux was integrated from the continuum-subtracted data; the results appear in Table~\ref{tab:spec}.
The integrated flux from the deep spectrum is within $3\sigma$ of the flux from the noisier shallow spectrum.  
 
To date, \cet\ is one of only two disks observed with \her\ that show \ion{C}{2} emission but no \ion{O}{1}.
The other is HD32297, a bright $\sim 30$~Myr-old debris disk system \citep{Donaldson:2013,Kalas:2005}.
HD32297 was previously known to contain atomic gas, but no molecular gas has yet been detected \citep{Redfield:2007a}.

\begin{deluxetable}{lrcccc}
\tablewidth{0pt}
\tablecolumns{6}
\tablecaption{Spectroscopy Results for 49~Ceti \label{tab:spec}}
\tablehead{ \colhead{Line} & \colhead{$\lambda_0$\tablenotemark{a}} &
\colhead{$F_\mathrm{cont}$\tablenotemark{b}} & \colhead{$F_\mathrm{int}$\tablenotemark{c}} & \colhead{$\sigma_{F_\mathrm{int}}$} & \colhead{$S/N$\tablenotemark{d}} \\
\colhead{ } & \colhead{(\um)} & \colhead{(Jy)} & \multicolumn{2}{c}{($\times 10^{-18} \ \mathrm{W} \: \mathrm{m}^{-2}$)} }
\startdata
\ion{O}{1} & 63.185  & $2.09 \pm 0.35$ & $< 11.05$ & 3.68 &  \\
CO J=36-35 & 72.843  & $1.95 \pm 0.32$ & $< 14.88$ & 4.96 &  \\
o-H$_2$O   & 78.741  & $1.90 \pm 0.31$ & $< 11.86$ & 3.95 & \\
CO J=29-28 & 90.163  & $1.88 \pm 0.32$ & $< 9.25$ & 3.08 & \\
\ion{O}{1} & 145.535 & $1.16 \pm 0.18$ & $< 6.16$ & 2.06 &  \\
\ion{C}{2} shallow & 157.741 & $1.13 \pm 0.20$ & 7.23 & 2.22 & $3.3 \sigma$ \\
\hspace*{3.4ex} deep &    & $0.98 \pm 0.13$ & 3.690 & 0.798 & $4.6 \sigma$ \\
o-H$_2$O   & 179.527 & $< 0.73$ & $< 12.17$ & 4.06 &  \\
DCO$^+$    & 189.570 & $< 1.37$ & $< 13.25$ & 4.42 & 
\enddata
\tablenotetext{a}{Line center rest wavelength.}
\tablenotetext{b}{Continuum flux at line center rest wavelength; upper limits are $3\sigma$. The flux errors include statistical and absolute flux calibration uncertainties. }
\tablenotetext{c}{Integrated emission line flux; upper limits are $3\sigma$.}
\tablenotetext{d}{Significance of emission line detection.}
\end{deluxetable} 

\subsection{Carbon Gas Mass \label{sub:gas_mass}}

With only one carbon emission line detected, we are not able to measure the excitation temperature of the gas from our data, which is needed to calculate a model-independent total mass of \ion{C}{2} ions from the integrated [\ion{C}{2}] 158~\um\ line flux.
However, we may determine a temperature-insensitive lower limit on the total mass, assuming the emission is optically thin.  
In this case, the mass is given by
\begin{equation} \label{eq:gas_mass}
M_\mathrm{CII} = \frac{4 \pi \, \lambda_0}{h c} \; \frac{F_{int} \: m \: d^2}{A_{ul} \: x_u} \; ,
\end{equation}
where $\lambda_0$ is the wavelength of the line, $F_{int}$ is the observed integrated emission line flux, $m$ is the mass of an atom, $d$ is the distance from the emitting region to the observer, $u$ and $l$ designate the upper and lower energy levels involved in the transition, $A_{ul}$ is the spontaneous transition probability\footnote{NIST Atomic Spectra Database: \url{http://www.nist.gov\newline/pml/data/asd.cfm}}, and $x_u$ is the fraction of atoms in the upper energy level.
Assuming local thermal equilibrium (LTE), $x_u$ is given by
\begin{equation}
x_u = \frac{ (2 \, J_u + 1) \; e^{-E_u/k T_\mathrm{ex}} }{Q_{T_\mathrm{ex}} } \: ,
\end{equation}
where $J_u$ is the angular momentum quantum number of the upper level, 
$E_u$ is the energy of the upper level, $T_\mathrm{ex}$ is the excitation temperature, and $Q_{T_\mathrm{ex}}$ is the partition function for the given excitation temperature. 

\begin{figure}[t!] \centering
\epsfig{file=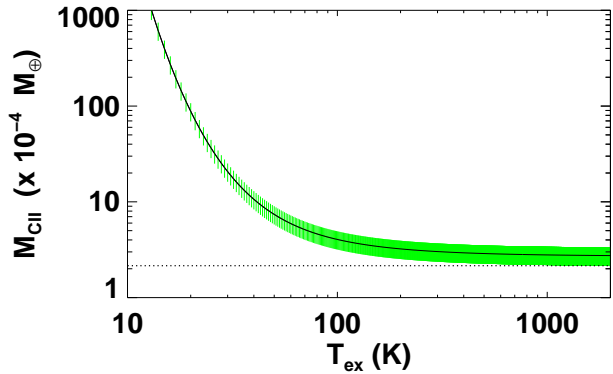, width=3.25in}
\caption{\cet\ \ion{C}{2} gas mass inferred from the 158~\um\ emission line flux as a function of excitation temperature. 
The $\pm 1 \sigma$ errors on the mass are plotted with green lines.
The horizontal dotted line shows the lower limit on the gas mass ($M_\mathrm{CII} \gtrsim 2.15 \times 10^{-4} \ M_\oplus$), valid for assumed excitation temperatures between 1~K and 2000~K. \label{fig:gas_mass} }
\end{figure}

Figure~\ref{fig:gas_mass} shows a plot of the \ion{C}{2} mass as a function of the assumed excitation temperature.
The lower limit on the mass is $M_\mathrm{CII} \gtrsim 2.15 \times 10^{-4} \ M_\oplus$, valid for $T_\mathrm{ex}$ between 1~K and 2000~K.
This value is close to the total CO mass calculated from the observed sub-mm CO emission \citep[$M_\mathrm{CO} = 2.2 \times 10^{-4} \ M_\oplus$;][]{Hughes:2008}.
Our upper limits on the CO 73 and 90~\um\ line fluxes do not provide useful checks on the CO abundance or excitation temperature.
These lines arise from very high energy levels ($E_u = 3471 \ \mathrm{and} \ 2240$~K, respectively) and therefore are insensitive tracers of the bulk of the CO gas.

To determine the total mass of carbon atoms in the \cet\ disk, we need to know the ionization balance in the gas.  This may be calculated with complex thermochemical disk models, further discussed in Section~5. Here we make a simple estimate of the mass lower limit by assuming a plausible ionization fraction, measured in an analogous environment.  The radiation and density characteristics of the \cet\ circumstellar environment are similar to those of $\beta$~Pic; both A stars are surrounded by optically thin disks containing a similar amount of dust, judging from the systems' fractional infrared luminosities. Both disks contain circumstellar gas, including CO \citep[although the $\beta$~Pic CO is not yet detected in sub-mm emission, only UV absorption;][]{Roberge:2000}.

The measured column densities of neutral and first-ionized carbon in the $\beta$~Pic gas are roughly equal, giving a carbon ionization fraction of $\sim 50\%$ \citep{Roberge:2006}, in agreement with the value predicted by photoionization calculations \citep{Fernandez:2006}. Therefore, the rough lower limit on the total mass of carbon atoms in the \cet\ gas is $M_\mathrm{C} \gtrsim 4.3 \times 10^{-4} \ M_\oplus$. 
If we further assume that the atomic gas has solar elemental abundances, we may estimate a lower limit on the total atomic gas mass by dividing the carbon mass limit by the solar carbon mass fraction \citep[0.288\%;][]{Lodders:2003}.
With all these assumptions (optically thin [\ion{C}{2}] emission, LTE, carbon $T_\mathrm{ex} < 2000~K$, 50\% ionization fraction, and solar elemental abundances), the rough lower limit on the total atomic gas mass is $M_\mathrm{a} \gtrsim 0.15 \ M_\oplus$.

Turning to the other undetected emission lines, we cannot at present use them to calculate model-independent upper limits on the masses of other gases, due to a lack of precise information on the gas excitation temperatures. 
Examination of Figure~\ref{fig:gas_mass} shows that the \ion{C}{2} mass vs.\ excitation temperature curve changes slowly at high temperatures but shoots up at low temperatures; this is a generic characteristic of such curves. 
It is easy to set a mass lower limit that is valid for a wide range of temperatures, since it's controlled by the highest temperature considered, where the curve is changing slowly. 
Setting a temperature-insensitive upper limit is difficult, since it's controlled by the lowest temperature considered, where the curve is changing rapidly. 

The situation is worse for the [\ion{O}{1}] 63~\um\ line, since the energy of the upper level ($E_u = 228$~K) is much higher than it is for the [\ion{C}{2}] 158~\um\ line ($E_u = 91$~K).
Therefore, the \ion{O}{1} mass upper limit is extremely sensitive to the range of temperatures considered.
For example, assuming LTE and changing the lowest excitation temperature considered from 10~K to 50~K decreases the \ion{O}{1} mass upper limit by 6 orders of magnitude. 
Furthermore, the large energy separation for the levels involved in the [\ion{O}{1}] 63~\um\ line means that the levels are unlikely to be thermalized and assuming LTE is not appropriate.

\section{SED Fitting \label{sec:sed} }

\begin{figure*}[ht] \centering \hspace*{-0.4in}
\epsfig{file=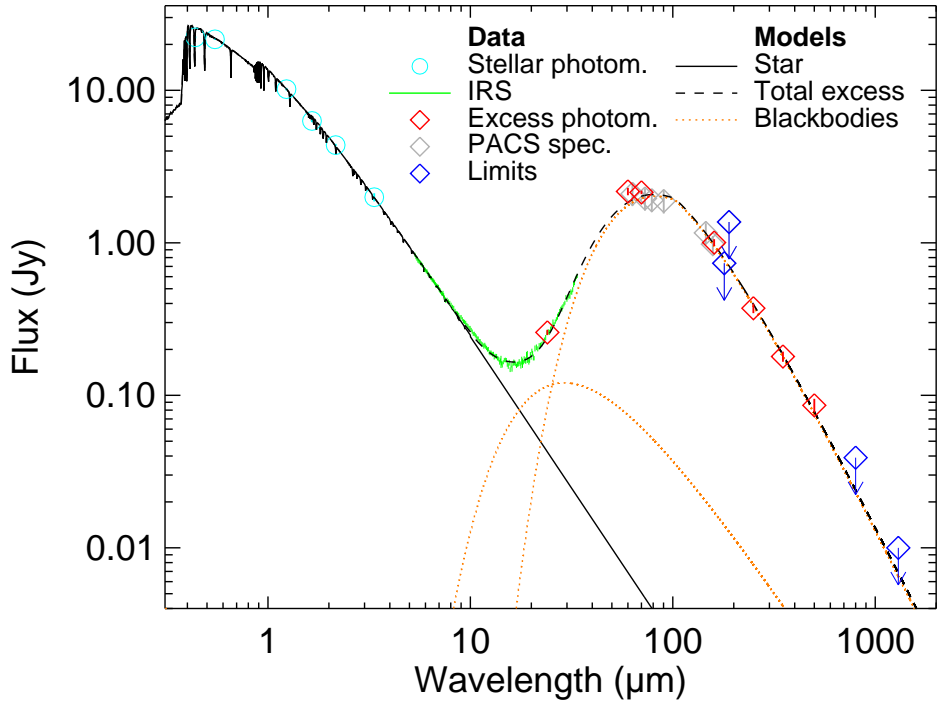, width=5.0in}
\caption{Spectral energy distribution for \cet\ with a two-component blackbody model for the dust excess emission.
Our best-fitting ATLAS9 stellar model (solid black line) was fit to the $B$ \& $V$ magnitudes from the Tycho-2 Catalog, the $J, H, \ \& \ K_s$ fluxes from 2MASS, and the WISE W1 flux (cyan circles).
The total model (black dashed line) is the sum of the best-fitting stellar model, a warm simple blackbody, and a cold modified blackbody.
It was fit to various photometry detections (red diamonds), continuum fluxes from PACS spectroscopy (grey diamonds), and the \si\ IRS spectrum (green line).
The individual blackbody components are plotted with orange dotted lines.
Some relevant flux upper limits are also shown (blue arrows).  \label{fig:sed} }
\end{figure*}

We have collected a wide range of unresolved photometry and spectroscopy of \cet; our compilation of continuum fluxes at wavelengths from the optical to the millimeter appears in Table~\ref{tab:all_phot}.
In Figure~\ref{fig:sed}, we show the most complete SED to date.
It includes our new far-IR/sub-mm continuum fluxes from \her;
the \textit{Spitzer} IRS spectrum; the \textit{Spitzer} MIPS 24~\um\ flux; and new near-IR photometry from \textit{WISE}. 
Some redundant fluxes given in Table~\ref{tab:all_phot} were not used in the SED.

Table~\ref{tab:all_phot} includes two unrefereed detections of \cet\ at sub-mm/mm wavelengths: one at 850~\um\ from JCMT/SCUBA \citep{Song:2004} and one at 1.2~mm from IRAM \citep{Bockelee-Morvan:1994}.
As noted by \citet{Hughes:2008}, the two fluxes cannot be simultaneously fitted with any dust excess model that decreases with increasing wavelength in the sub-mm/mm.
It is not obvious which flux value is more accurate.
In the absence of other sub-mm fluxes, \citet{Hughes:2008} conservatively adopted the lower flux value from SCUBA \citep{Song:2004}.
The new SPIRE fluxes allow us to reasonably constrain the SED fitting without relying on the ambiguous SCUBA and IRAM fluxes.
Therefore, we avoided making an arbitrary choice between them by excluding both points from our new SED.
We eagerly await sensitive new sub-mm/mm observations of \cet\ with ALMA.

%
%
\begin{deluxetable*}{lccr@{$\; \pm \;$}lp{2.2in}}[ht]
\tablecolumns{6}
\tablewidth{0pt}
\tablecaption{Unresolved Continuum Fluxes for \cet\ \label{tab:all_phot}}
\tablehead{ \colhead{Source} & \colhead{Band} & \colhead{Wave.} & \multicolumn{2}{c}{Flux} & \colhead{Reference} \\
\colhead{ } & \colhead{or Mode} & \colhead{(\um)} & \multicolumn{2}{c}{(Jy)} & \colhead{ } }
\startdata
Tycho-2 & Johnson $B$ & 0.44 & 22.26 & 0.35 & \citet{Hog:2000} \\
Tycho-2 & Johnson $V$ & 0.55 & 21.60 & 0.20 & \citet{Hog:2000} \\
2MASS   & $J$         & 1.24 & 10.18 & 0.19 
& 2MASS All-Sky Catalog of Point Sources\\
2MASS   & $H$         & 1.65 &  6.30 & 0.13 
& 2MASS All-Sky Catalog of Point Sources\\
2MASS   & $K_s$       & 2.16 &  4.373 & 0.081 
& 2MASS All-Sky Catalog of Point Sources\\
WISE\tablenotemark{a} & W1          & 3.35 &  1.99 & 0.12 & \citet{Wright:2010}\\
WISE\tablenotemark{a} & W2          & 4.60 &  1.294 & 0.049
& \citet{Wright:2010}\\
WISE\tablenotemark{a} & W3          & 11.56 & 0.211 & 0.021 & \citet{Wright:2010}\\
IRAS\tablenotemark{b}    &          & 12.00 & 0.228 & 0.034 & IRAS Faint Source Catalog, v2.0\\
Keck$/$MIRLIN &    & 12.50 & 0.200 & 0.026 & \citet{Wahhaj:2007}\\
Keck$/$MIRLIN &    & 17.90 & 0.186 & 0.025 & \citet{Wahhaj:2007}\\
WISE\tablenotemark{a} & W4         & 22.09 & 0.238 & 0.024 & \citet{Wright:2010}\\
Spitzer$/$MIPS\tablenotemark{c} &   & 24.00 & 0.259 & 0.010 & This work \\
IRAS\tablenotemark{b}    &    & 25.00 & 0.312 & 0.042 & IRAS Faint Source Catalog, v2.0\\
IRAS\tablenotemark{b}    &    & 60.00 & 2.17 & 0.12 & IRAS Faint Source Catalog, v2.0\\
IRAS\tablenotemark{b}    &    & 100.00 & 1.88 & 0.21 & IRAS Faint Source Catalog, v2.0\\
Herschel$/$PACS\tablenotemark{d} & Spec. & 63.19 & 2.09 & 0.35 & This work\\
Herschel$/$PACS\tablenotemark{d} & Phot. & 70.00  & 2.142 & 0.058 & This work\\
Herschel$/$PACS\tablenotemark{d} & Spec. & 72.84 & 1.95 & 0.32  & This work\\
Herschel$/$PACS\tablenotemark{d} & Spec. & 78.74 & 1.90 & 0.31  & This work\\
Herschel$/$PACS\tablenotemark{d} & Spec. & 90.16 & 1.88 & 0.32  & This work\\
Herschel$/$PACS\tablenotemark{d} & Spec. & 145.54 & 1.16 & 0.18 & This work\\
Herschel$/$PACS\tablenotemark{d} & Spec. & 157.68 & 0.98 & 0.13 & This work\\
Herschel$/$PACS\tablenotemark{d} & Phot. & 160.00 & 1.004 & 0.053 & This work\\
Herschel$/$PACS\tablenotemark{d} & Spec. & 179.53 & \multicolumn{2}{c}{$\leq 0.73 \ (3 \sigma)$} & This work\\
Herschel$/$PACS\tablenotemark{d} & Spec. & 189.57 & \multicolumn{2}{c}{$\leq 1.37 \ (3 \sigma)$} & This work\\
Herschel$/$SPIRE\tablenotemark{d} & Phot. & 250.00 & 0.372 & 0.027 & This work\\
Herschel$/$SPIRE\tablenotemark{d} & Phot. & 350.00 & 0.180 & 0.014 & This work\\
Herschel$/$SPIRE\tablenotemark{d} & Phot. & 500.00 & 0.086 & 0.009 & This work\\
JCMT    &          & 800.00 & \multicolumn{2}{c}{$\leq 0.039 \ (3 \sigma)$} & \citet{Zuckerman:1993}\\
JCMT$/$SCUBA &     & 850.00 & 0.0082 & 0.0019 & \citet{Song:2004}\\
IRAM    &          & 1200.00 & 0.0127 & 0.0028 & \citet{Bockelee-Morvan:1994}\\
SMA     &	   & 1300.00 & \multicolumn{2}{c}{$\leq 0.01 \ 
(3 \sigma)$\tablenotemark{e}} & \citet{Hughes:2008}
\enddata
\tablenotetext{a}{WISE W1, W2, and W3 fluxes are color-corrected for a $\nu^{2}$ spectrum. The W4 flux is color-corrected for a 100~K blackbody.
The WISE errors include statistical and absolute calibration uncertainties added in quadrature (see  \url{http://wise2.ipac.caltech.edu/docs/release/allsky/expsup/sec4\_4h.html}).}
\tablenotetext{b}{IRAS fluxes are color-corrected (see \url{http://lambda.gsfc.nasa.gov/product/iras/colorcorr.cfm}).}
\tablenotetext{c}{For the Spizter MIPS 24~\um\ flux, the error is the 4\% absolute calibration uncertainty.}
\tablenotetext{d}{For all Herschel data, the flux errors are the statistical and  calibration uncertainties added in quadrature. The SPIRE fluxes are color-corrected for a $\nu^2$ spectrum (see \url{http://herschel.esac.esa.int/Docs/SPIRE/html/spire\_om.html}).}
\tablenotetext{e}{Upper limit on total continuum flux, assuming the dust emission is spread over 4.8~SMA beams (beam size = $1\farcs0 \times 1\farcs2$).  This is the size of the CO $J=2-1$ emission region in the SMA images \citep{Hughes:2008}.}
\end{deluxetable*}

\subsection{Stellar Model \label{sub:star} }

To determine the expected stellar flux at all wavelengths, we compared photospheric models to the \cet\ SED shortward of about 8~\um, where the dust excess starts to become noticeable.
A careful analysis of the \cet\ stellar parameters appears in \citet{Montesinos:2009}.
We were motivated to revisit that work by the existence of important information not available or utilized at the time: specifically, the independent stellar age estimate \citep[$\sim 40$~Myr;][]{Zuckerman:2012} and the {\it Spitzer} IRS spectrum. 
We found that while the photospheric model computed using the best stellar parameters in \citet{Montesinos:2009} fit the optical and near-IR fluxes well, it appeared slightly too bright when compared with the short wavelength portion of the IRS spectrum.

Therefore, we fit ATLAS9 solar-metallicity photospheric models\footnote{Downloaded from F.~Castelli's public database at \url{http://wwwuser.oat.ts.astro.it/castelli/grids.html.}} to the optical and near-IR data \citep{Castelli:2004}.
The data used were the $B$- and $V$-band fluxes from the Tycho-2 Catalog, the $J$, $H$, and $K_s$ fluxes from 2MASS, and the color-corrected WISE Band~1 flux. No interstellar reddening was applied to the models.
The best-fitting model had an effective temperature $T_\mathrm{eff} = 9000$~K and also matched the short wavelength portion of the IRS spectrum well.
This temperature is typical of an A1V star but slightly cooler than the 9500~K effective temperature in \citet{Montesinos:2009}. 
However, this fitting procedure could not provide a useful constraint on the stellar gravity.

Using the new effective temperature and age estimate, we placed the star on a $\log g - \log T_{\rm eff}$ H-R diagram, forcing it to lie precisely on the 40-Myr isochrone.  We used the evolutionary tracks and isochrones with solar metallicity from the Yonsei-Yale group \citep{Yi:2001}. The stellar gravity indicated by the age constraint is $\log g=4.3$, the corresponding stellar mass is 2 $M_\odot$, and the stellar luminosity is 15.5 $L_\odot$. A photospheric model with $T_{\rm eff}=9000$ K and $\log g=4.3$ was built by interpolating two ATLAS9 solar-metallicity photospheric models with $\log g=$4.0 and 4.5. No extinction was applied to the model, which was then normalized to the IRS spectrum at 6~\um. The final stellar model is shown with a black solid line in Figure~\ref{fig:sed}.

A strong check of the internal consistency of the set of stellar parameters consists of using the stellar luminosity derived from the age constraint ($L_* = 15.5~L_\odot$) and the total photospheric flux ($F_*$, computed by integrating the final stellar model) to recover the observed distance.
Using the simple expression $L_* = 4 \pi \, d^2 \, F_*$, the derived distance is 56~pc, in excellent agreement with the \emph{Hipparcos} distance of 59~pc.  
A lower gravity would imply a larger luminosity and hence a much greater distance; as an example, $T_{\rm eff}=9000$ K and $\log g=4.0$ in the $\log L_*/L_\odot - \log T_{\rm eff}$ H-R diagram corresponds to a luminosity of 34.67~$L_\odot$ (using PMS tracks) or 36.31~$L_\odot$ (using post-MS tracks). These luminosities give distances of $83 - 85$~pc, inconsistent with the \emph{Hipparcos} distance.

\subsection{Modified Blackbody Dust Models \label{sub:bb}}

We began our analysis of the \cet\ dust excess emission with a very simple model, i.e.\ the sum of the best stellar model and a single-temperature modified blackbody.  
Starting from equation~6 in \citet{Beckwith:1990}, an optically thin modified blackbody has the form
\begin{equation} 
F_\nu(T) \: = \: B_\nu(T) \: \left( 1 - e^{-\tau_\nu} \right) \: \Omega \;
\simeq \; B_\nu(T) \: \tau_\nu \: \Omega \: ,
\end{equation}
where $B_\nu(T)$ is the Planck function for temperature $T$, $\tau_\nu$ is the optical depth, and $\Omega$ is the solid angle of the emitting area (effectively a constant of proportionality).
The optical depth is proportional to the dust mass opacity.
At mid-IR to mm wavelengths, mass opacities for expected debris disk grains have a power-law dependence on wavelength \citep[see Figure~7 in][]{Lebreton:2012}.
At shorter wavelengths, the mass opacity for simple astronomical silicate grains abruptly becomes nearly independent of wavelength. Therefore, our formulation for the optical depth is  
\begin{equation}
\begin{array}{rclr}
\tau_\nu & = & \tau_0 & \lambda < \lambda_0 \\
 & = & \tau_0 \; \left( \frac{\lambda}{\lambda_0} \right)^{- \beta}
& \ \ \ \ \ \lambda > \lambda_0   \label{eq:tau_nu}
\end{array}
\end{equation}
where $\tau_0$ is the optical depth at reference wavelength $\lambda_0$ and $\beta$ is the emissivity power-law index. 

This simple model was least-squares fit to the \cet\ SED, ignoring flux upper limits. 
We found that a single-temperature blackbody model could not simultaneously fit the mid-IR emission seen with \si\ and the far-IR emission peak.
The parameters for the best dust model that fits the far-IR data appear in Table~\ref{tab:sed}.
The fractional infrared luminosity (\lir), which is the ratio of the blackbody model integrated over all frequencies to the similarly integrated stellar model, is proportional to $\tau_0$.
The reduced $\chi^2$ value is extremely large due to the model's failure to also match the numerous points in the IRS spectrum.
That failure indicates the disk has dust grains at a wide range of temperatures, which can be interpreted as a warm inner dust component and cold outer material. \citet{Wahhaj:2007} came to the same conclusion and fit the \cet\ SED with a two-component dust disk.


\begin{deluxetable}{lccc}
\tablecaption{Parameters of the Best Blackbody Dust Models \label{tab:sed}}
\tablecolumns{4}
\tablehead{\multicolumn{1}{l}{Parameters} & \colhead{Single-Temperature} & \multicolumn{2}{c}{Two-Temperature}} 
\startdata
$T$ (K) & $83 \pm 9$ & $62 \pm 1$ & $175 \pm 3$ \\
$\lambda_0$ (\um) & $156 \pm 35$ & $102 \pm 12$ & \nodata \\
$\beta$ & $0.6 \pm 0.1$ & $0.7 \pm 0.1$ & \nodata \\
\lir\ $(\times 10^{-4})$ & $11 \pm 2$ & $7.1 \pm 0.2$ & $1.26 \pm 0.03$\\
Reduced $\chi^2$ & $4 \times 10^{6}$ & \multicolumn{2}{c}{1.18}
\enddata
\end{deluxetable}

We therefore made a total SED model consisting of the best stellar model, a warm simple blackbody, and a cold modified blackbody.
This model was also least-squares fit to the \cet\ SED, ignoring flux upper limits as before.
The best-fitting two-temperature model is over-plotted in Figure~\ref{fig:sed}, along with the two individual blackbody components; 
the model parameters appear in Table~\ref{tab:sed}.
The reduced $\chi^2$ value is greatly improved compared to the single-temperature model.

The temperature of the warm component ($175 \pm 3$~K) is higher than the temperature found by fitting a single-temperature simple blackbody to the IRS spectrum alone \citep[$118 \pm 6$~K;][]{Chen:2006}.
This is presumably because we are now accounting for the flux in the IRS spectrum coming from the cold component.
The emissivity power-law index found for the cold component ($\beta = 0.7 \pm 0.1$) is smaller that the values typically seen in the interstellar medium \citep[$\beta \approx 2$; e.g.][]{Andrews:2005}.
This is often seen in protoplanetary and debris disk SEDs and it indicates that the emitting grains are larger than the typical sub-micron interstellar grains \citep[e.g.][]{Su:2006}.

Assuming the dust grains are in radiative equilibrium with the central star, we estimated the minimum characteristic radius of each dust component. Starting from equation~1 in \citet{Beckwith:1990}, we adopted simple blackbody grains and used the bolometric stellar flux from our best stellar model. More realistic grains will be further from the star than simple blackbody grains at the same temperature, so the distances calculated are lower limits on the characteristic dust radii.  The characteristic radius of the warm component is $\gtrsim 11$~AU, while that of the cold component is $\gtrsim 84$~AU.

\section{Gas Modeling \label{sec:gas}}

There are two possible explanations for the gas content of the \cet\ disk.  As mentioned in the Introduction, the most common previous scenario interprets the CO as the last remnant of the primordial gas left over from star formation \citep[e.g.][]{Hughes:2008}.
In the first subsection, we describe our attempts to fit the \cet\ SED and gas observations using a primordial-like disk model.
The next subsection considers the second explanation, that the \cet\ gas is secondary material coming from the destruction of planetary material.

\subsection{Primordial Gas Scenario}

We began a detailed investigation of the \cet\ gas emission using an advanced disk modeling code, ProDiMo, which self-consistently determines the temperature, density, and chemical structure of combined gas and dust disks \citep{Woitke:2009, Kamp:2010, Thi:2011}.
ProDiMo calculates the heating and cooling of both gas and dust, and takes into account 960 chemical reactions involving 71 species.
Photochemistry, H$_2$ formation on grain surfaces, and cosmic-ray heating are all included.

As a first step, we considered the \cet\ disk model from \citet{Hughes:2008} that successfully reproduced the basic features of the spatially and spectrally resolved sub-mm CO emission, as well as the dust emission from the outer disk (the inner disk was ignored).
In that work, steady-state chemistry was adopted and solar elemental abundances assumed.
Some key features of the model were 1) a lack of gas within 40~AU of the star, 2) an outer disk radius of 200~AU, and 3) a total gas mass of $13~M_\oplus$.

In the original \citet{Hughes:2008} model, the gas-to-dust ratio had to be set to the unusually high value of 500.
This was likely caused by necessary simplifying assumptions about the dust properties: well-mixed gas and dust, a uniform dust composition (astronomical silicates), and a single large grain size.
Here we retained the first two assumptions but used a Dohnanyi power-law grain size distribution \citep[$dn \propto s^{-3.5} ds$;][]{Dohnanyi:1969}.
We then calculated several models that reproduced the observed sub-mm CO emission while varying other disk parameters (e.g.\ gas-to-dust ratios from 300 to 65). 
While it was possible to find models that matched the sub-mm CO emission and the upper limit on [\ion{O}{1}] 63~\um\ emission, they under-predicted the [\ion{C}{2}] 158~\um\ emission by factors of $5 - 23$.

In sum, simple adjustments to the \citet{Hughes:2008} model were unsuccessful at simultaneously reproducing the sub-mm CO emission, the \ion{C}{2} emission, and the lack of \ion{O}{1} emission.
For example, reducing the oxygen abundance in an attempt to decrease the \ion{O}{1} emission led to less CO formation and reduced CO cooling, which resulted in a warmer gas disk.  Paradoxically, this had the effect of increasing the \ion{O}{1} emission rather than decreasing it. 
The only adjustment that showed some promise involved increasing the carbon abundance over the solar value.
Further work on detailed modeling of the \cet\ disk with ProDiMo is underway. 
For now, it appears that the observations cannot be reproduced with a ``primordial-like'' protoplanetary disk model.

\subsection{Secondary Gas Scenario}

We therefore consider whether the gas could also be consistent with a secondary source from planetesimals.
If so, its composition could shed light on the make-up of the young parent planetesimals -- outcomes of the planet formation process and the building blocks of full-sized planets.
Questions that can be addressed include whether the parent planetesimals are rocky or icy and whether there are any abundance anomalies \citep[e.g.][]{Roberge:2006, Xie:2012}.

In this scenario, the presence of relatively abundant CO (for a debris disk) suggests that this gas comes from icy material, more comet-like than asteroid-like \citep[as proposed in][]{Zuckerman:2012}.
There are several possible mechanisms for its production, including outgassing of comet-like bodies, photodesorption of ice-coated grains, and grain-grain collisions \citep{Lagrange:1998, Chen:2007, Czechowski:2007}.
%
%
How large a mass of comets would be needed to supply the observed mass of CO?
In equilibrium, the CO production rate will match the loss rate.
Assuming the CO is in a low-density environment, chemical reactions can be ignored and the primary loss mechanism is photodissociation. 

Therefore, the CO loss rate is 
\begin{equation}
\left( \frac{d n_\mathrm{CO}}{dt} \right)_{\mathrm{loss}} = - \: k \; n_\mathrm{CO} = - \: k \left( \frac{X_\mathrm{CO}}{V} \right) 
\end{equation} 
where $n_\mathrm{CO}$ is the CO number volume density, $k$ is the photodissociation rate, $X_\mathrm{CO}$ is the total number of CO molecules, $V$ is the total volume of the gas.
Given the low \lir\ value, shielding of the CO from dissociating UV radiation by dust grains should be negligible.
Shielding by H$_2$ is hard to assess, although most hydrogen in comets is locked in H$_2$O rather than H$_2$ \citep[e.g.][]{Mumma:2011}.
For now, we ignore the competing effects of shielding and the stellar UV field, and consider only the interstellar field.
The unshielded photodissociation rate for CO in a Draine interstellar UV field is $k = 2.6 \times 10^{-10} \ \mathrm{s}^{-1}$, giving an unshielded CO lifetime  $\tau_\mathrm{CO} = 1/k \approx 120$~yrs \citep{Visser:2009}.
The CO production rate is
\begin{equation}
\left(\frac{d n_\mathrm{CO}}{dt} \right)_{\mathrm{prod.}} = + \: r_\mathrm{CO} \; \frac{d n_\mathrm{H_2O}}{dt} 
 = + \: \frac{r_\mathrm{CO}}{V} \: \frac{d \, X_\mathrm{H_2O}}{dt} \: ,
\end{equation}
where $r_\mathrm{CO}$ is the fractional abundance of CO relative to water in comets, $d n_\mathrm{H_2O} / dt$ is the water production rate, and
$X_\mathrm{H_2O}$ is the total number of $\mathrm{H_2O}$ molecules.
Observations of Solar System comets show $r_\mathrm{CO}$ values ranging from about $0.4\%$ to $30\%$ \citep{Mumma:2011}.

Equating the production and loss rates shows that equilibrium is achieved for a water mass loss rate of 
\begin{equation}
\frac{d \, M_\mathrm{H_2O}}{dt} = \frac{k}{r_\mathrm{CO}} \;  \frac{m_\mathrm{H_2O}}{m_\mathrm{CO}} \; M_\mathrm{CO} \: ,
\end{equation}
where $M_\mathrm{H_2O}$ is the total mass of H$_2$O, $m_\mathrm{H_2O}$ is the mass of a H$_2$O molecule, $m_\mathrm{CO}$ is the mass of a CO molecule, and $M_\mathrm{CO}$ is the total mass of CO \citep[$2.2 \times 10^{-4} \ M_\oplus$;][]{Hughes:2008}.
Therefore, the water mass loss rate needed to produce the observed CO is 
\begin{equation}
\frac{d \, M_\mathrm{H_2O}}{dt} \: \approx \: 10^{-13} \ \mathrm{to} \ 10^{-11} \ M_\oplus \ \mathrm{s}^{-1} \: .
\end{equation}
Since comets are mostly water, we can adopt these values as approximate comet mass loss rates.
Collisional modeling of the Solar System's Kuiper Belt predicts an initial mass of $\sim 60 \ M_\oplus$ \citep{CampoBagatin:2012}.
Scaling this value up by the larger stellar mass of \cet\ ($\approx 2 \ M_\odot$) gives an initial planetesimal mass of $\sim 120 \ M_\oplus$. 
With the loss rates given above, it would take between roughly 0.4~Myr and 32~Myr to exhaust the total initial planetesimal mass.
Since the star is likely to be $\sim 40$~Myr old \citep{Zuckerman:2012}, this analysis would suggest that if the CO is being produced from comets, it is a relatively short-lived phenomenon.

\section{Discussion \label{sec:discussion}}

The dissipation of abundant primordial material left over from star formation sets crucial constraints on the formation of planetary systems. 
On one hand, the removal of the gas limits the time available for formation of gas giant planets. 
On the other hand, the presence of a modest amount of gas during the later stages of formation can help damp the inclinations and eccentricities of terrestrial planets \citep[e.g.][]{Nagasawa:2005}.
Therefore, understanding the speed of gas dissipation and the mechanisms by which it occurs is vitally important for informing planet formation theories.

Primordial dust lifetimes are fairly well-constrained by observations to be $< 10$~Myr, although there is a large dispersion in values for individual stars of nearly the same age and uncertainties remain about the effects of stellar mass, binarity, and star-forming environment \citep[e.g.][]{Haisch:2001, Andrews:2005, Carpenter:2006, Cieza:2009, Luhman:2010}.
Determining gas lifetimes, one of the primary goals of the \her\ GASPS project, is far harder.
While arguments can be made that gas and dust dissipate on roughly similar timescales \citep{Roberge:2011}, there is significant uncertainty about the co-evolution of gas and dust \citep[e.g.][]{Pascucci:2009}.

From this perspective, \cet\ might be a nearly unique disk system that is just finishing dissipation of its primordial gas.  
With gas emission reminiscent of a low-mass protoplanetary disk, \cet\ does look to be in some sort of intermediate state.
As pointed out in \citet{Hughes:2008}, the presence of an inner dust disk that lacks significant molecular gas could be consistent with disk dissipation from the inside out through photoevaporation, if the inner disk grains are large enough not to be entrained in a photoevaporative flow \citep[e.g.][]{Alexander:2006}.
With the previous age estimate of $\sim 9$~Myr \citep{Montesinos:2009}, \cet\ would have a relatively long but not implausible primordial gas lifetime.
However, the new 40~Myr age \citep{Zuckerman:2012} requires an anomalously long lifetime for the observed CO to be primordial gas.
Our preliminary disk modeling with ProDiMo casts further doubt on the \cet\ gas being primordial, since we have great difficulty fitting all the gas observations with a low-mass but otherwise normal protoplanetary disk model.

Looking at \cet's dust properties, it greatly resembles a young debris disk.
The low fractional dust luminosity \citep[$\lesssim 1\%$ of a typical Herbig Ae disk value;][]{Meeus:2012} and the lack of 10~\um\ silicate emission indicating few small grains in the warm inner disk are both highly characteristic of debris disks \citep[e.g.][]{Chen:2006}.
All of the estimated ages for \cet\ could place it in a late stage of terrestrial planet formation \citep[e.g.][]{Kenyon:2006}.
Furthermore, recent work has shown that many debris disk SEDs are best fit by two-temperature models, like \cet, suggesting radially separated inner and outer dust belts \citep{Morales:2011}.  

The \cet\ gas may also be secondary debris material coming from destruction of planetesimals. 
Our simple calculations of gas production from comet-like ices indicate that the amount of material required to produce the observed CO is not implausibly large.
Unfortunately, we are not able to say at this time if the lower limit on the \ion{C}{2} mass and non-detection of \ion{O}{1} emission are also consistent with the comet-evaporation scenario.
First, if the gas is coming from planetesimals, then the parent species for the bulk of the carbon gas is not necessarily CO.
In the better-studied case of $\beta$~Pic, balancing \ion{C}{1} production by photodissociation of CO with \ion{C}{1} loss by photoionization suggests that only about 2\% of the total carbon gas comes from dissociation of CO \citep{Roberge:2000, Roberge:2006}.
Since the primary atomic gas production mechanism is likely to be photodesorption of grain surfaces or grain-grain collisions, many other materials could supply carbon gas to the disk (e.g.\ amorphous carbon).
This makes interpreting the lower limit on the \cet\ \ion{C}{2} mass difficult.
Second, while we expect oxygen gas as an end-product of water ice evaporation, we are not able to turn the observed \ion{O}{1} flux upper limit into a useful oxygen mass upper limit (see Section~\ref{sub:gas_mass}).

However, recent monitoring of \ion{Ca}{2} lines in optical spectra of \cet\ show absorption features that are variable in both strength and velocity shift \citep{Montgomery:2012}.
Similar features are also seen in spectra of $\beta$~Pic and are attributed to so-called ``falling evaporating bodies'' (FEBs), which are star-grazing planetesimals passing through the line of sight to the central star \citep[e.g.][]{Beust:1990}.
Therefore, each absorption feature is effectively a transit of an exo-comet or exo-asteroid. 
Taken together, the difficulty fitting the \cet\ observations with a primordial disk model, the possibility of producing the observed CO from comet-like material, and the detection of gas-producing planetesimals all indicate that \cet\ is a close analog of the better-understood $\beta$~Pic debris disk.
Tighter constraints on the gas-to-dust ratios in both systems, through more accurate determination of dust masses and measurements of additional gas species, would help confirm their similarity and allow studies of the effect of the gas on grain dynamics.

\section{Concluding Remarks}

The \her\ far-IR imaging and spectroscopy of \cet\ has provided an extremely rich dataset for better characterizing both the dust and gas. 
Our primary findings are as follows.
\begin{enumerate}
\item In our 70~\um\ image, dust continuum emission from the outer disk is spatially resolved for the first time.
The disk orientation agrees with that determined from spatially and spectrally resolved sub-mm CO emission, suggesting the gas and dust are co-spatial.
The deconvolved image does not show a central clearing in the dust; 
the Gaussian HWHM along the disk major axis is $\sim 200$~AU. 
\item The [\ion{C}{2}] 158~\um\ emission line was detected at the $3 \sigma$ level in a preliminary shallow spectrum and at the $5 \sigma$ level in a deep follow-up spectrum. 
No other emission lines were detected, including [\ion{O}{1}] 63~\um.
The integrated \ion{C}{2} line flux is $(3.690 \pm 0.798) \times 10^{-18} \ \mathrm{W \ m^{-2}}$, giving an ionized carbon gas mass $M_\mathrm{CII} \gtrsim 2.15 \times 10^{-4} \ M_\oplus$.
With various assumptions, this suggests a lower limit on the total atomic gas mass of $M_\mathrm{a} \gtrsim 0.15 \ M_\oplus$.
\item Simple modeling of the new SED confirms the two-component structure of the disk.  We find blackbody temperatures of 175~K for the inner dust and 62~K for the outer dust, implying characteristic radii of $\gtrsim 11$~AU and $\gtrsim 84$~AU.
\item Preliminary thermochemical modeling of the combined gas and dust disk indicates that a ``primordial-like'' protoplanetary disk model cannot simultaneously reproduce all the gas observations. 
The gas may instead be secondary material coming from the destruction of comet-like ices.
\end{enumerate}

The next step in our work on the \cet\ system is fitting the SED with an advanced dust disk modeling code \citep[GRaTeR;][]{Augereau:1999}.
This code includes realistic grain properties and can take into account the powerful constraints provided by resolved imaging, as was done in our modeling of the HD181327 debris disk \citep{Lebreton:2012}.
The new dust disk model will provide a more accurate dust mass and can be input to ProDiMo for further exploration of the disk parameter space.
We will also experiment with ProDiMo runs in ``debris disk mode,'' adopting time-dependent chemistry and non-solar starting compositions.
Through these and other studies, we hope to resolve the nature of the interesting \cet\ disk.


\acknowledgments


%
\her\ is an ESA space observatory with science instruments provided by European-led Principal Investigator consortia and with important participation from NASA.
Support for this work was provided by the NASA \her\ Science Center through an award issued by JPL/Caltech. A.~Roberge also acknowledges support by the Goddard Center for Astrobiology, part of the NASA Astrobiology Institute. \mbox{J.-C.} Augereau thanks the CNES-PNP for financial support. C.~Eiroa, G.~Meeus, and B.~Montesinos were partly supported by Spanish grant
AYA 2011-26202.

{\it Facilities:} \facility{Herschel (PACS, SPIRE)},  \facility{Spitzer (IRS, MIPS)}, \facility{Hipparcos}, \facility{CTIO:2MASS}, \facility{WISE}, 
\facility{IRAS}, \facility{JCMT}, \facility{SMA}.






\end{document}